\documentclass[reprint,prx,twocolumn,showpacs,superscriptaddress,aps,longbibliography]{revtex4-2}

\usepackage[colorlinks=true,citecolor=blue, linkcolor=blue, urlcolor=blue]{hyperref} 
\usepackage{graphicx}
\usepackage{bm}
\usepackage{amsmath}
\usepackage{amssymb}
\usepackage{svg}

\DeclareGraphicsExtensions{.png,.jpg,.eps,.svg}
\usepackage{xcolor}

\newcommand{\beq}{\begin{equation}}
\newcommand{\eeq}{\end{equation}}

\usepackage{mleftright} 

\newcommand{\figref}[1]{\mbox{Fig.~\ref{#1}}}

\newcommand{\secref}[1]{\mbox{Section~\ref{#1}}}

\renewcommand{\eqref}[1]{\mbox{Eq.~(\ref{#1})}}

\newcommand{\figpanel}[2]{Fig.~\hyperref[#1]{\ref*{#1}(#2)}}
\newcommand{\figpanels}[3]{Fig.~\hyperref[#1]{\ref*{#1}(#2)--(#3)}}
\newcommand{\figpanelNoPrefix}[2]{\hyperref[#1]{\ref*{#1}(#2)}}

\newcommand{\ket}[1]{|#1 \rangle}
\newcommand{\bra}[1]{\langle #1|}

\newcommand{\Tr}[1]{\text{Tr}#1}

\usepackage{siunitx}

\begin{document}

\author{Walter Rieck}
\affiliation{Department of Microtechnology and Nanoscience, Chalmers University of Technology, 41296 Gothenburg, Sweden}
\affiliation{Department of Physics, Stockholm University, 10691 Stockholm, Sweden}

\author{Ariadna Soro}
\affiliation{Department of Microtechnology and Nanoscience, Chalmers University of Technology, 41296 Gothenburg, Sweden}

\author{Anton Frisk Kockum}
\email{anton.frisk.kockum@chalmers.se}
\affiliation{Department of Microtechnology and Nanoscience, Chalmers University of Technology, 41296 Gothenburg, Sweden}

\author{Guangze Chen}
\email{guangze@chalmers.se}
\affiliation{Department of Microtechnology and Nanoscience, Chalmers University of Technology, 41296 Gothenburg, Sweden}

\title{Controlled-Z gates with giant atoms in structured waveguides}

\begin{abstract}

Giant atoms are quantum emitters coupled to waveguides at multiple, spatially separated points, enabling interference effects that fundamentally change their light--matter interactions. A notable consequence of the interference is the emergence of decoherence-free interaction (DFI), which allows coherent excitation exchange between giant atoms via the waveguide without radiative loss. Leveraging DFI offers a promising route to implementing two-qubit quantum gates without the need for additional resources, positioning giant atoms as a versatile platform for scalable universal quantum simulators. However, existing work has focused primarily on continuous, Markovian waveguides; in structured waveguides, where non-Markovian effects become significant, only iSWAP gates have been explored. To address this gap, we introduce and analyze a protocol for implementing controlled-Z (CZ) gates with giant atoms in structured waveguides. We first show that while a minimal two-point coupling scheme supports DFI, it also exhibits strong non-Markovian effects that substantially degrade gate fidelity. To overcome this limitation, we propose an extended design featuring a third coupling point. This configuration suppresses non-Markovian effects and enables CZ gates with fidelities up to \qty{97.7}{\percent} (assuming typical values for experimental imperfections). Our results broaden the accessible gate set for giant atoms in structured waveguides to include both iSWAP and CZ gates, advancing these systems as a pathway toward universal quantum simulators operating in non-Markovian environments.

\end{abstract}

\date{\today}

\maketitle


\section{Introduction}

Universal quantum simulators are powerful tools for exploring complex open quantum many-body dynamics~\cite{Georgescu2014, Altman2021, Fauseweh2024}. Yet, most existing quantum simulators are optimized for implementing only specific two-qubit gates~\cite{Wintersperger2023, Fauseweh2024, Strohm2024}. Although this can be sufficient for realizing a universal gate set together with single-qubit gates, it is generally desirable to have more flexibility regarding which two-qubit gates one can implement, e.g., to reduce the depth of quantum circuits. However, enabling such flexibility typically requires additional coupling elements~\cite{PhysRevApplied.10.034050, PhysRevLett.123.210501, PhysRevA.102.062408, Abrams2020, Ganzhorn2020, Lacroix2020,  PhysRevApplied.16.024050, Sete2021, Sung2021, Rigetti,     Zhang2024, Chen2025a, PhysRevApplied.23.024059, Krizan2025, lvb9-pfr3} or complicated drive schemes~\cite{Wei2022, Nguyen2024, PRXQuantum.5.020326, Evered2023, PhysRevLett.125.120504, PRXQuantum.5.020338} that limit the feasibility of building large-scale quantum simulators. 

Here, we explore how an extended set of two-qubit gates can be realized using giant atoms (GAs)~\cite{Kockum2021, Gustafsson2014, Kockum2014, Kockum2018, Kannan2020}, which have recently emerged as a versatile platform for scalable universal quantum simulators that can handle open quantum systems~\cite{Chen2025, Yang2025, chen2025scalablequantumsimulatorextended, Chen2025b}. Unlike conventional small atoms that interact locally, GAs couple to waveguides at multiple, spatially separated points. This nonlocal coupling results in interference-driven light–matter dynamics, giving rise to qualitatively new phenomena such as frequency-dependent decay~\cite{Kockum2014, Kannan2020, Vadiraj2021, Wang2022a}, directional emission~\cite{Chen2022a, Joshi2023, Wang2024, Jouanny2025a}, tunable interaction rates~\cite{Kockum2018, Kannan2020, Wang2022a} between distant GAs, and much more~\cite{Kockum2021, Guo2017, Gonzalez-Tudela2019, Guo2020, Guimond2020, Ask2020, Cilluffo2020, Wang2021, Du2021, Soro2022, Wang2022, Du2022, Du2022a, Terradas-Brianso2022, Soro2023, Du2023, Ingelsten2024, Leonforte2024, Roccati2024, Gong2024, Du2025, Du2025a, Levy-Yeyati2025, Li2025, Rieck2025, Manenti2017, Satzinger2018, Bienfait2019, Andersson2019, Bienfait2020, Andersson2020, Hu2024, Almanakly2025, Xiao2025}.
Among the most striking consequences of this nonlocal coupling is the possibility of realizing decoherence-free interactions (DFIs), where two or more atoms exchange excitations coherently while completely suppressing radiative decay through destructive interference between coupling points~\cite{Kockum2018, Kannan2020}. These DFIs provide a natural mechanism for entangling gates without the need for additional parametric couplers, making GAs a versatile platform for quantum simulators~\cite{Chen2025, Yang2025, chen2025scalablequantumsimulatorextended, Chen2025b}.

Previous studies have often focused on GAs coupled to continuous waveguides, whose dispersion can be approximated as linear in the frequency regime where the atoms operate. In these setups, the Markovian approximation is typically valid and analytical treatments are tractable~\cite{Chen2025, chen2025scalablequantumsimulatorextended, Chen2025b}. Additionally, in such setups, iSWAP gates have been experimentally demonstrated using superconducting qubits coupled to a microwave waveguide~\cite{Kannan2020}. 

In contrast to continuous waveguides, structured waveguides~\cite{Calajo2016, RevModPhys.90.031002, PhysRevLett.119.143602}, such as coupled-cavity arrays~\cite{PhysRevX.12.031036, Jouanny2025, PhysRevLett.134.133601, Jouanny2025a} or photonic crystals~\cite{Liu2016, PhysRevX.9.011021, PhysRevA.99.052126}, have nonlinear dispersions with band edges and gaps in the frequency regime where the atoms operate. Exploring GAs coupled to structured waveguides thus opens up the possibility of exploiting band-edge effects, engineered dispersion, and strong non-Markovian effects~\cite{Soro2023, PhysRevA.107.023716, Longhi2020, Zhao2020, Wang2021, Longhi2021, Yu2021, Vega2021, Wang2022, Xiao2022, Cheng2022, Zhang2022SC, Du2023a, Du2023b, Bag2023, Jia2024, Gao2024}. Yet, since the Markovian approximation does not hold, such studies require keeping track of the whole system's dynamics and have thus mainly focused on single-excitation dynamics, which only support iSWAP gates~\cite{Soro2023}. Establishing universal quantum simulators in GAs coupled to structured waveguides would benefit from additional entangling gates such as the controlled-Z (CZ) gate, which remains to be explored as it relies on dynamics in the two-excitation sector.

In this article, we propose and analyze a protocol for implementing a CZ gate between two GAs coupled to a structured waveguide. We begin by studying the minimal two-point coupling configuration, which supports DFI, but suffers from non-Markovian effects that strongly degrade the DFI. We then demonstrate that introducing a third coupling point shifts the decoherence-free frequencies closer to the center of the photonic band of the structured waveguide, thereby mitigating these effects. For experimentally realistic parameters in superconducting qubits, we show that this design yields CZ gate fidelities up to \qty{97.7}{\percent}. These results equip GAs coupled to structured waveguides with an extended gate set involving both iSWAP and CZ gates, pushing these platforms forward toward universal quantum simulators in non-Markovian environments.

The rest of this article is organized as follows. In \secref{sec2}, we introduce the model of two three-level GAs coupled to a structured waveguide. In \secref{sec3}, we analyze the CZ gate protocol, beginning with the minimal two-point configuration and then extending to the improved three-point scheme. We further evaluate the resulting gate fidelity and discuss experimental feasibility. Finally, in \secref{sec4}, we summarize the main findings and outline their broader implications.


\section{Model and method}
\label{sec2}

\subsection{System Hamiltonian}

\begin{figure}
    \centering
    \includegraphics[width=1\linewidth]{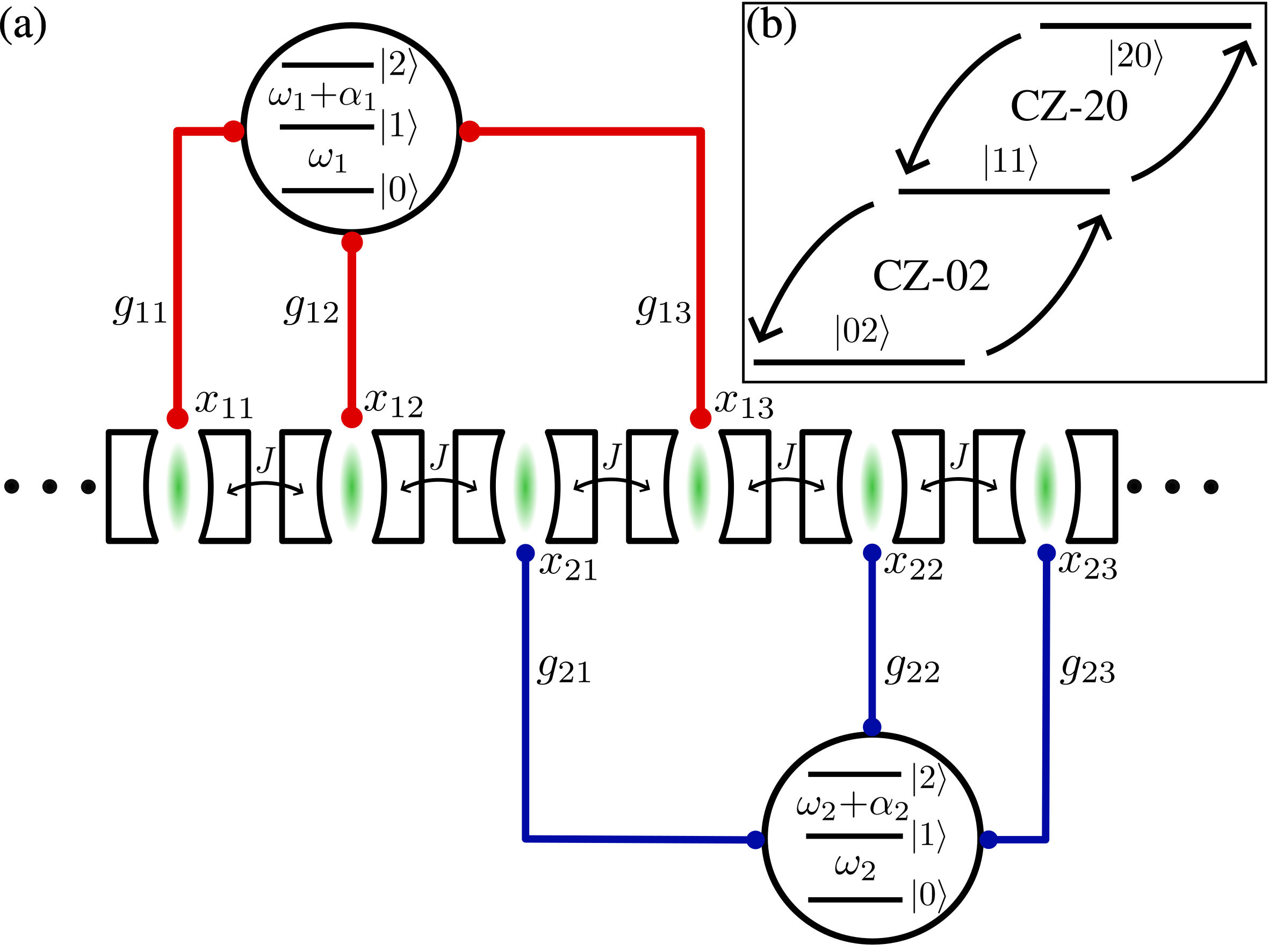}
    \caption{Model and gate scheme.
    (a) Schematic of two giant atoms (GAs) coupled to a one-dimensional array of coupled cavities with nearest-neighbor hopping rate $J$. Each atom $m$ is connected with coupling strength $g_{mj}$ to the array at site $x_{mj}$. The atoms have distinct transition frequencies $\omega_{1(2)}$ and anharmonicities $\alpha_{1(2)}$. 
    (b) The CZ gate in GAs can be implemented via the pathway CZ-02 (CZ-20), which coherently transfers the population of the $\ket{11}$ state to the $\ket{02}$ ($\ket{20}$) state and then brings it back, accumulating a $\pi$ phase shift.}
    \label{fig:setup}
\end{figure}

We consider a system of two GAs coupled to a one-dimensional structured waveguide, as shown in \figpanel{fig:setup}{a}. The total Hamiltonian for this system is given by
\beq \label{eq:Hamiltonian}
H = H_A + H_B + H_{\rm int},
\eeq
where $H_A$ describes the bare GAs, $H_B$ describes the bare bath (i.e., the waveguide), and $H_{\text{int}}$ describes the interaction between the atoms and the bath. Since the CZ gate dynamics in our scheme involve the $\ket{2}$ level of the qubits [see \figpanel{fig:setup}{b}], we model each atom as a three-level system ($\hbar = 1$ here and in the rest of the article):
\beq \label{eq_HA}
H_A = \sum_{m=1}^2 \mleft[\mleft(2\omega_m + \alpha_m\mright)\frac{\sigma_m^{(2), +}\sigma_m^{(2), -}}{2} + \omega_m\sigma_m^{(1), +}\sigma_m^{(1), -}\mright],
\eeq
where $\omega_m$ denotes the transition frequency from the ground state to the first excited state for atom $m$, $\alpha_m$ denotes the anharmonicity of atom $m$, and the ladder operators are defined as
\beq \label{eq_sigma}
\sigma_m^{(j), +}\ket{j-1}_m=\sqrt{j}\ket{j}_m,\quad \sigma_m^{(j), -}\ket{j}_m=\sqrt{j}\ket{j-1}_m .
\eeq

We model the structured bath as a one-dimensional array of $N$ coupled cavities, with the center of the photonic band at the frequency $\omega_c$:
\beq \label{eq_HB}
H_B = \sum_{n=1}^{N} \omega_c a_n^\dag a_n -J \sum_{n=1}^{N-1} \mleft(a_n a_{n+1}^{\dag} + \text{H.c.} \mright) ,
\eeq
where $J$ is the coupling rate between adjacent cavities, $a_n$ and $a_n^\dagger$ are the annihilation and creation operators, respectively, for photons in cavity $n$, and H.c.~denotes the Hermitian conjugate. This Hamiltonian yields the dispersion relation
\beq \label{eq:DispRel}
\omega(k) = -2J\cos{k}+\omega_c,
\eeq
where $k$ is the wave number, resulting in a cosine band of width $4J$. As $\omega_c$ is a constant, we can go to a frame rotating at that frequency in \eqref{eq:Hamiltonian} such that the first term in \eqref{eq_HB} is zero and we redefine $\omega_m=\omega_m-\omega_c$ in \eqref{eq_HA}. 

Under the rotating-wave approximation, the interaction Hamiltonian between the atoms and the bath takes the form
\beq
H_{\text{int}} = g_{mj}\sum_{m=1}^{2}\sum_{j=1}^{p_m} \mleft[ \sigma_m^{(1), +}a_{x_{mj}}+\sigma_m^{(2), +}a_{x_{mj}}+ \text{H.c.}\ \mright],
\eeq
where $x_{mj}$ denotes the position of the $j$th coupling point of atom $m$ and $g_{mj}$ the corresponding coupling strength. Each atom is coupled to $p_m$ cavities in total.

\subsection{Numerical approach to dynamics}

Solving the full Hamiltonian $H$ in~\eqref{eq:Hamiltonian} is computationally challenging, as it acts on an infinite-dimensional Hilbert space. However, since $H$ conserves the total excitation number of the atom–bath system, it can be projected onto subspaces with a fixed number of excitations. In particular, the dynamics of the iSWAP gate only involve single-excitation states, and the problem can therefore be solved by restricting the Hamiltonian to the single-excitation subspace. In contrast, the dynamics of the CZ gate involve two excitations, going beyond the single-excitation regime that is typically studied in giant-atom systems~\cite{Soro2023}. As a result, we must numerically simulate the system in the two-excitation subspace. This subspace is spanned by the states 
\beq \label{eq7}
\begin{aligned}
    \ket{20}_A\otimes\ket{0}_B ,
    & \qquad \ket{11}_A\otimes\ket{0}_B , \\
    \ket{02}_A\otimes\ket{0}_B , & \qquad \ket{10}_A\otimes\ket{j}_B , \\
    \ket{01}_A\otimes\ket{j}_B , & \qquad \ket{00}_A\otimes\ket{jk}_B ,    
\end{aligned}
\eeq
where $\ket{n_1,n_2}_A$ denotes the state of the two atoms with excitation numbers $n_{1,2} \in \{0, 1, 2\}$, $\ket{0}_B$ is the vacuum state of the waveguide, $\ket{j}_B$ represents a single photon in cavity $j$, and $\ket{jk}_B$ corresponds to two photons in the waveguide---one in cavity $j$ and one in cavity $k$ ($j = k$ covers the case of two photons in one cavity). Since the photons are bosons, we impose the symmetry condition $\ket{jk}_B = \ket{kj}_B$. Using this basis, we project the full many-body description from the Hamiltonian in \eqref{eq:Hamiltonian} onto the two-excitation subspace and simulate the system dynamics accordingly.

\subsection{Computing gate fidelity}
\label{sec:ComputingGateFidelity}

The average gate fidelity of the CZ gate ${F}_{\text{CZ}}$ is related to the process fidelity $\mathcal{F}_{\text{CZ}}$ by
\beq
{F}_{\text{CZ}} = \frac{d \mathcal{F}_{\text{CZ}}+1}{d+1},
\eeq
where $d=4$ is the Hilbert-space dimension of the computational subspace for the two-qubit CZ gate~\cite{Gilchrist2005}.  

To evaluate $\mathcal{F}_{\text{CZ}}$, we first describe the state of the total system as
\beq
\ket{\psi} = \sum_{m,n} \alpha_{mn} \ket{m}_A \otimes \ket{n}_B,
\eeq
where $\ket{m}_A$ and $\ket{n}_B$ denote atomic and bath states, respectively [see \eqref{eq7}]. The corresponding density matrix is
\beq
\rho_{\text{tot}} = \sum_{m,n}\sum_{m',n'} \alpha_{mn}\alpha^*_{m'n'} 
\ket{m}_A \bra{m'}_A \otimes \ket{n}_B \bra{n'}_B.
\eeq
Tracing out the bath degrees of freedom gives the reduced atomic density matrix
\beq
\rho_A = \Tr_B \rho_{\text{tot}} = \sum_{m,m',n} \alpha_{mn}\alpha^*_{m'n}\ket{m}_A\bra{m'}_A.
\eeq
Since the computational subspace only involves the atomic states $\{\ket{11}, \ket{10}, \ket{01}, \ket{00}\}$, we can compute $\rho_A$ as the product of a matrix $M$ with its conjugate transpose,
\beq
M_{mn} =
\begin{cases}
    \alpha_{mn}, & \text{if } \ket{m}_A\ket{n}_B \in \text{basis}, \\
    0, & \text{otherwise},
\end{cases}
\quad \rho_A = M M^\dagger.
\eeq
Using this reduced density matrix, a process $\mathcal{E}$ can be represented by its Choi matrix~\cite{Choi1975}
\beq
\Phi = \sum_{m,n=11,10,01,00} \ket{m}\bra{n} \otimes \mathcal{E}(\ket{m}\bra{n}),
\eeq
where $\ket{m}$, $\ket{n}$ runs over the computational basis states and $\mathcal{E}(\ket{m}\bra{n})$ denotes the reduced density matrix after evolving the corresponding input state. The process fidelity is then defined as~\cite{Gilchrist2005}
\beq
\mathcal{F}_{\text{CZ}} = \left[ \Tr \left( \sqrt{ \sqrt{\Phi} \, \Phi_{\text{true}} \, \sqrt{\Phi} } \right) \right]^2,
\eeq
where $\Phi_{\text{true}}$ is the Choi matrix of the ideal CZ gate.

\begin{figure}[t!]
    \centering
    \includegraphics[width=1\linewidth]{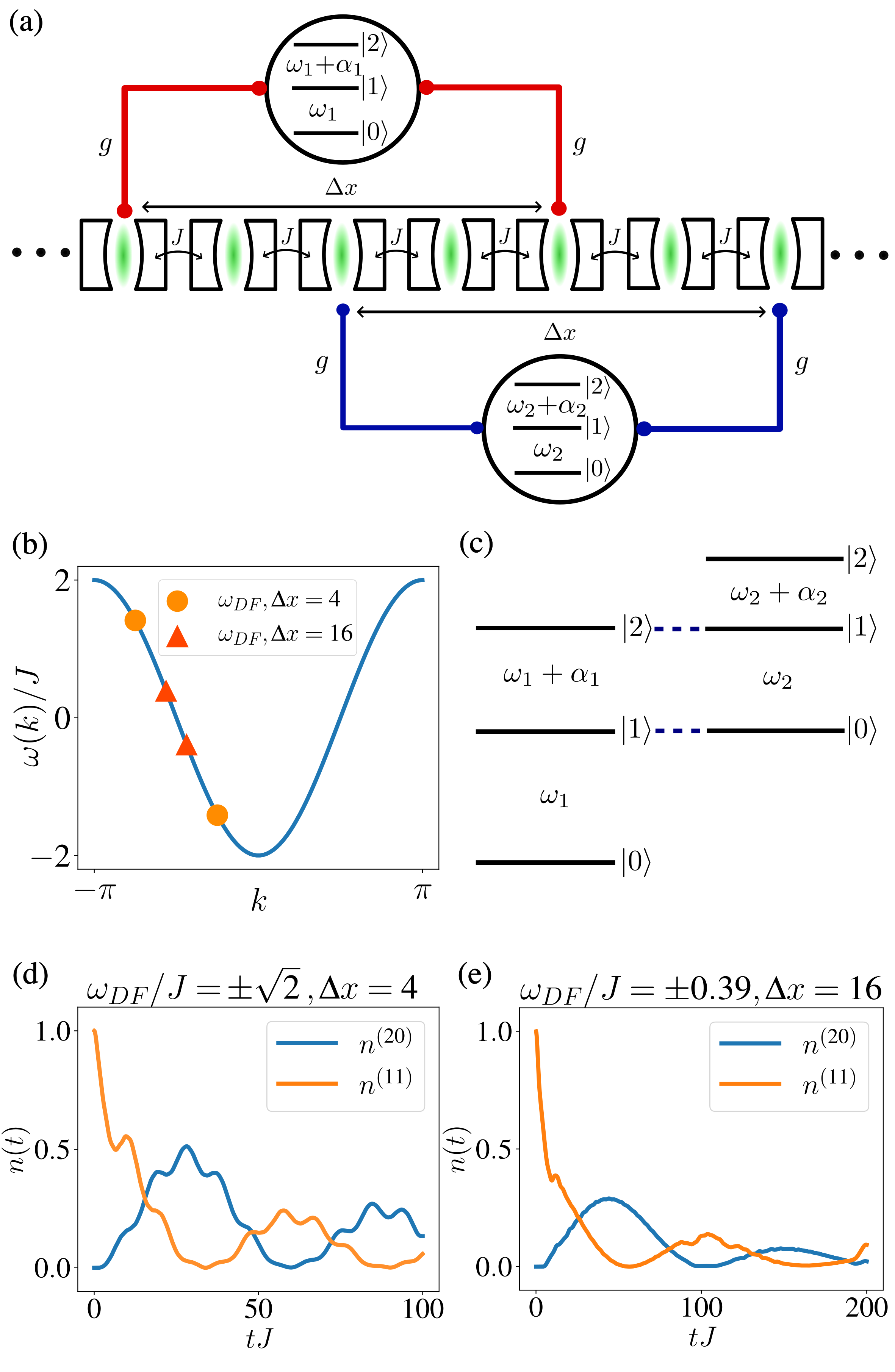}
    \caption{Setup and results for performing a CZ gate between GAs with two coupling points each.
    (a) Schematic of the GA-based setup with two coupling points to perform a CZ gate.
    (b) Bath dispersion relation and the decoherence-free frequencies for $\Delta x = 4$ ($\omega_{\rm DF}/J=\pm\sqrt{2}$) and $\Delta x = 16$ ($\omega_{\rm DF}/J=\pm0.39$) shown with circles and triangles, respectively.
    (c) Energy-level diagram of the two GAs for settings enabling the CZ gate.
    (d) Dynamics in the decoherence-free regime, with parameters $N=100$, $g/J = 0.175$, $\omega_1/J = \sqrt{2}$, $\omega_2/J = -\sqrt{2}$, $\alpha_1/J = -2\sqrt{2}$, and $\alpha_2/J = -3$. 
    (e) Same as in (d), but with increased separation $\Delta x = 16$ and parameters $\omega_1/J = 0.39$, $\omega_2/J = -0.39$, $\alpha_1/J = -0.78$, and $\alpha_2/J = -1$. 
 }
    \label{fig2}
\end{figure}


\section{Controlled-Z gate} 
\label{sec3}

We now investigate how a CZ gate can be implemented using GAs coupled to a structured waveguide. The CZ gate applies a phase shift of $\pi$ to the $\ket{11}$ state of a two-qubit system while leaving all other basis states unchanged. In our system, this operation can be achieved by coherently transferring the population of the $\ket{11}$ state to $\ket{02}$ or $\ket{20}$ and then bringing it back~\cite{Strauch2003}; see \figpanel{fig:setup}{b}. This process requires a resonant transition between these states. In our setup, we realize this resonance by setting $\omega_2 = \omega_1 + \alpha_1$, which enables a resonant coupling between $\ket{11}$ and $\ket{20}$. 

To prevent radiative decay during the evolution, we require that transitions at both $\omega_1$ and $\omega_1 + \alpha_1$ must be decoherence-free. Thus, implementing the CZ gate requires at least two decoherence-free frequencies within the photonic band. For a GA with coupling points ${x_j}$ and coupling strengths ${g_j}$, the decoherence-free condition arises from destructive interference~\cite{Kockum2014}:
\beq \label{eq8}
\sum_j g_j \exp(i k_{\text{DF}} x_j) = 0 ,
\eeq
where $k_{\text{DF}}$ is a wave number that fulfills this equation.


\subsection{Minimal two-point implementation} 
\label{sec3.1}

The most straightforward configuration that enables fulfilling all the above conditions involves two coupling points separated by $\Delta x$ and uniform coupling strength, as shown in \figpanel{fig2}{a}.
In this case, \eqref{eq8} reduces to
\beq
1 + e^{i k_{\text{DF}} \Delta x} = 0 \quad \Rightarrow \quad k_{\text{DF}} = \frac{\pi + 2n\pi}{\Delta x}, \quad n \in \mathbb{N}.
\eeq
Using the dispersion relation from \eqref{eq:DispRel}, we obtain the corresponding decoherence-free frequencies:
\beq
\omega_{\text{DF}} = -2J \cos \left( \frac{\pi + 2n\pi}{\Delta x} \right), \quad n \in \mathbb{N}.
\eeq
To obtain two distinct DF frequencies within the band, we require $\Delta x \geq 3$. However, for $\Delta x = 3$, one of the DF frequencies is located at the band edge ($\omega_{\text{DF}} = 2J$), where strong non-Markovian effects dominate~\cite{Soro2023}. We therefore choose $\Delta x = 4$, yielding DF frequencies at $\omega_{\text{DF}} = \pm \sqrt{2} J$, as shown in \figpanel{fig2}{b}.

To perform the CZ gate, the level spacings of the two GAs need to satisfy $\omega_1+\alpha_1=\omega_2$ [\figpanel{fig2}{c}] and that both $\omega_1$ and $\omega_2$ are decoherence-free. These conditions are satisfied for $\Delta x=4$ with 
$\omega_1/J = \sqrt{2}$, $\omega_2/J = -\sqrt{2}$, and $\alpha_1/J = -2\sqrt{2}$. The parameter $\alpha_2$ does not qualitatively influence the dynamics; we set $\alpha_2/J = -3$ to have it similar to the anharmonicity of atom 1.

The dynamics with the system configured thus and initialized in $\ket{11}_A$ is shown in \figpanel{fig2}{d}. The populations of the $\ket{20}$ and $\ket{11}$ states are computed as 
\begin{align}
n^{(20)}(t) &= \frac{1}{2} \bra{\psi(t)} \sigma_1^{(2),+} \sigma_1^{(1),+}\sigma_1^{(1),-} \sigma_1^{(2),-} \ket{\psi(t)} , \\
n^{(11)}(t) &= \bra{\psi(t)} \sigma_2^{(1),+} \sigma_1^{(1),+}\sigma_1^{(1),-} \sigma_2^{(1),-} \ket{\psi(t)} , 
\end{align}
where $\ket{\psi(t)}$ is the state at time $t$. Although this setup formally satisfies the DF condition, the DF frequencies lie far from the band center, leading to strong non-Markovian effects~\cite{Soro2023}. As a result, population decays rapidly, the DFI breaks down, and the two-point configuration fails to support a high-fidelity CZ gate. 

A natural way to address this issue is to increase the separation $\Delta x$, which shifts the DF frequencies closer to the band center. For example, $\Delta x = 16$ yields $\omega_{\text{DF}}/J = \pm 0.39$, as shown in \figpanel{fig2}{b}. However, as shown in \figpanel{fig2}{e}, this strategy does not improve performance: the larger separation introduces a significant time delay for photons traveling between coupling points, enhancing non-Markovian effects leading to significant population loss into the waveguide.

\begin{figure}[t!]
    \centering
    \includegraphics[width=1\linewidth]{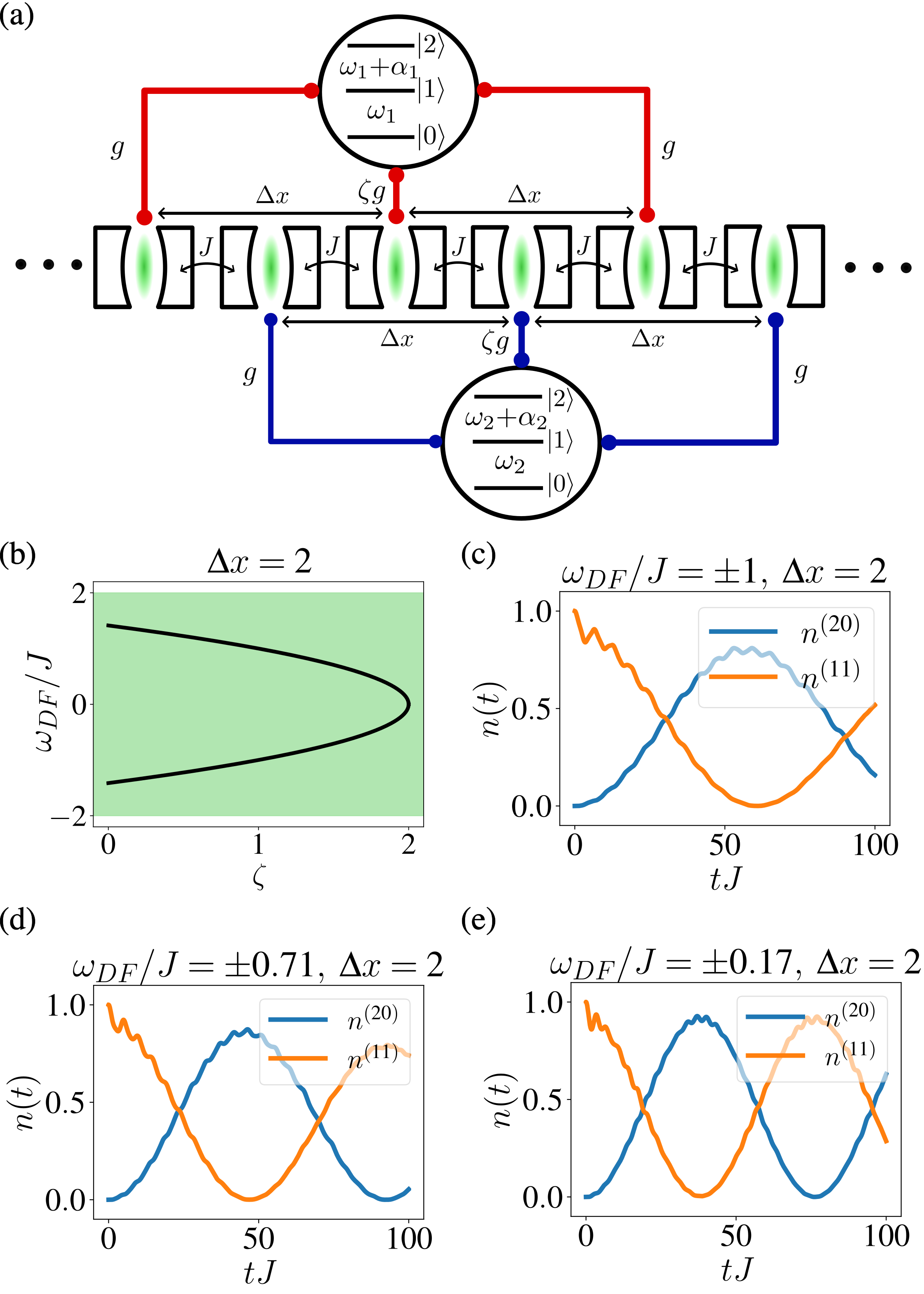}
    \caption{Setup and results for performing a CZ gate between GAs with three coupling points each.
    (a) Schematic of the GA-based setup with three coupling points to perform a CZ gate, with $\Delta x = 2$ between adjacent coupling points. Compared to \figpanel{fig2}{a}, a third, central coupling point with coupling strength $\zeta g$ is added. This configuration yields two decoherence-free (DF) frequencies for $0 \leq \zeta < 2$.  
    (b) Decoherence-free frequencies as a function of $\zeta$, showing that the two frequencies merge at $\zeta = 2$ with the band of propagating frequencies between $[-2, 2]$ shown in light green.
    (c) Dynamics in the DFI regime for $N=100$, $\Delta x = 2$, $g/J = 0.1$, and $\zeta = 1$, yielding DF frequencies $\omega_\text{DF}/J = \pm 1$. The frequencies and anharmonicities of the GAs are chosen to fit in the CZ-gate configuration: $\omega_1/J = 1$, $\omega_2/J = -0.98$, $\alpha_1/J = -2$, and $\alpha_2/J = -1.52$.  
    (d) Same as (c), but with $\zeta = 1.5$, giving DF frequencies $\omega_\text{DF}/J = \pm 0.71$. Here, we take parameters $\omega_1/J = 0.71$, $\omega_2/J = -0.69$, $\alpha_1/J = -1.42$, and $\alpha_2/J = -1.31$.  
    (e) Same as (c) and (d), but with $\zeta = 1.97$, resulting in DF frequencies $\omega_\text{DF}/J = \pm 0.17$. Here, we take parameters $\omega_1/J = 0.17$, $\omega_2/J = -0.17$, $\alpha_1/J = -0.34$, and $\alpha_2/J = -0.67$.
}
    \label{fig3}
\end{figure}


\subsection{Improvement with a three-point implementation} 
\label{sec3.2}

To overcome the limitations of the two-point configuration, we introduce a third coupling point placed in the middle between the original two, with a relative coupling strength $\zeta$, as shown in \figpanel{fig3}{a}. In this setup, the decoherence-free condition in \eqref{eq8} is modified to
\beq
1 + \zeta e^{i k_{\text{DF}} \Delta x} + e^{2 i k_{\text{DF}} \Delta x} = 0.
\eeq
The corresponding decoherence-free wavevectors are
\beq
k_{\text{DF}} = \frac{ \arccos \mleft( -\frac{\zeta}{2} \mright) + 2n\pi }{ \Delta x }, \quad n \in \mathbb{N},
\eeq
with associated frequencies
\beq
\omega_{\text{DF}} = -2J \cos \mleft[ \frac{ \arccos \mleft( -\frac{\zeta}{2} \mright) + 2n\pi }{ \Delta x } \mright], \quad n \in \mathbb{N}.
\eeq

The dependence of $\omega_{\text{DF}}$ on $\zeta$ for $\Delta x = 2$ is shown in \figpanel{fig3}{b}. As $\zeta$ increases, the DF frequencies shift progressively toward the center of the photonic band. At $\zeta = 2$, the two solutions merge at $\omega_{\text{DF}} = 0$. 

The dynamics for this configuration are illustrated in \figpanels{fig3}{c}{e} for three successively increasing values of $\zeta$ matched to the decoherence-free conditions required for the CZ gate. As $\zeta$ increases, the quality of the DFI improves: after a full cycle, a larger fraction of the population returns to $\ket{11}$, which is essential for high-fidelity CZ gate operation. This enhancement stems from the suppression of non-Markovian effects, since the relevant DF frequencies are brought closer to the band center. 

We note that the $\ket{1}\to\ket{0}$ transition of atom 2 is also off-resonantly coupled to the $\ket{1}\to\ket{0}$ transition in atom 1. As a consequence, to obtain a coherent population exchange between the atoms, $\omega_2$ needs to deviate slightly from $\omega_1+\alpha_1$ to take into account the Lamb shift~\cite{PhysRev.72.241} resulting from this coupling; hence the parameter choices in \figpanels{fig3}{c}{d}. We also note that the gate dynamics become faster for larger $\zeta$, as a consequence of the increased effective coupling between the atom and the waveguide.


\begin{figure}[t!]
    \centering
    \includegraphics[width=1\linewidth]{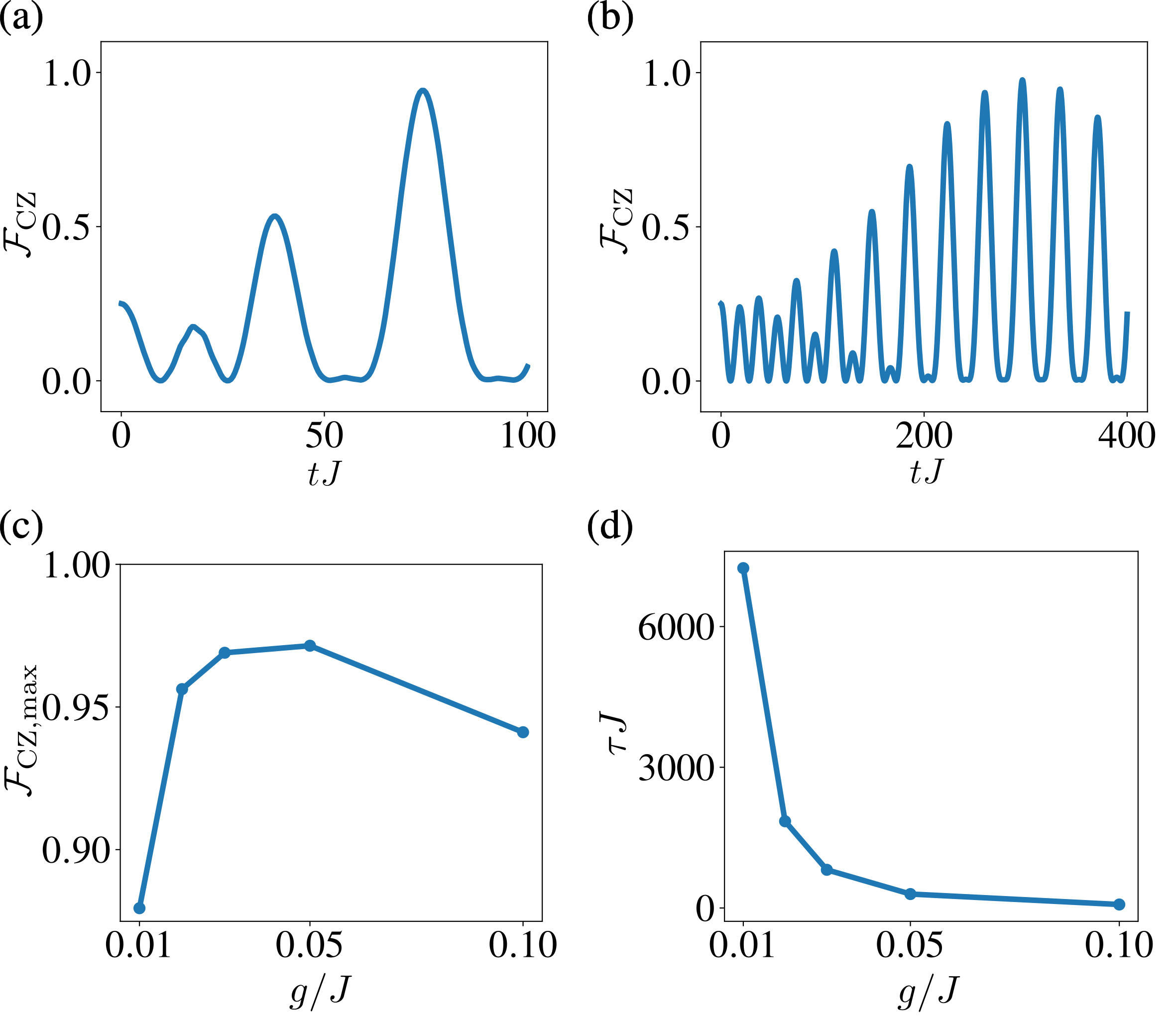}
    \caption{Process fidelity $\mathcal{F}_{\text{CZ}}$ of CZ gates performed with the setup in \figpanel{fig3}{a}.
    (a) Process fidelity as a function of time $t$ for the same parameters as in \figpanel{fig3}{e}.
    (b) Same as (a), but with a smaller $g=0.05J$. (c) The best process fidelity of CZ gates, $\mathcal{F}_{\text{CZ,max}}$, for different values of $g$, considering realistic qubit and cavity decay rates of $\Gamma_q=1.6\cdot10^{-5}J$ and $\Gamma_c=8\cdot10^{-5}J$. 
    (d) The gate time $\tau$ required for the best CZ gate in (c).
}
    \label{fig_fidelity}
\end{figure}

\subsection{Gate fidelity} 
\label{sec3.3}

Having analyzed the improved setup for performing a CZ gate with GAs in \figref{fig3}, we now compute the corresponding gate fidelity, as described in \secref{sec:ComputingGateFidelity}. The process fidelity $\mathcal{F}_{\text{CZ}}$ during the time evolution is shown in \figpanel{fig_fidelity}{a} for the parameters from \figpanel{fig3}{a}. A maximum process fidelity of $\qty{94.2}{\percent}$ is achieved at $tJ = 74.0$, corresponding to the time at which the population is transferred from $\ket{11}$ to $\ket{20}$ and back. The corresponding average gate fidelity is ${F}_{\text{CZ}} \approx \qty{95.4}{\percent}$. For ideal qubits and cavities, the fidelity can be further improved by reducing the coupling strength $g$. For example, choosing $g = 0.05J$ increases the process fidelity to $\mathcal{F}_{\text{CZ}} \approx \qty{97.6}{\percent}$, reached at $tJ = 296.8$~[\figpanel{fig_fidelity}{b}].

In realistic scenarios, where qubits and cavities exhibit intrinsic decay, a comprehensive optimization of the coupling strength $g$ would require solving the full master equation. This significantly increases the computational cost, as the Liouvillian superoperator scales with the square of the Hilbert-space dimension~\footnote{For $N=100$, the Hamiltonian Hilbert-space dimension is approximately $5500$, implying a Liouvillian dimension of $5500^2 = 3.025 \times 10^7$, beyond the reach of current exact-diagonalization techniques.}. To make the problem tractable, we instead analyze decay effects using the following effective non-Hermitian Hamiltonian, obtained by neglecting quantum jumps and collective decay processes:
\beq \label{eq_Heff}
\begin{aligned} 
H_{\rm eff}=&H-\frac{i\Gamma_q}{2}\sum_{m=1}^2\mleft[\sigma_m^{(2), +}\sigma_m^{(2), -} + \sigma_m^{(1), +}\sigma_m^{(1), -}\mright]
\\&-\frac{i\Gamma_c}{2}\sum_{n=1}^Na^\dag_na_n.
\end{aligned}
\eeq
Here, $H$ denotes the ideal system Hamiltonian defined in \eqref{eq:Hamiltonian} and $\Gamma_{q(c)}$ is the qubit (cavity) decay rate, assumed identical for all qubits (cavities). The higher decay rate of the $\ket{2}$ level compared to $\ket{1}$ is incorporated through the definition of $\sigma_m^{(2), \pm}$ in \eqref{eq_sigma}.

We now evaluate the CZ gate fidelity for experimentally relevant parameters in superconducting circuits, where GAs have already been demonstrated~\cite{Kannan2020, Vadiraj2021, Joshi2023, Hu2024, Jouanny2025a}. In particular, transmon qubits~\cite{Koch2007} coupled to microwave resonators provide a promising platform for implementing our proposal. In state-of-the-art devices, the transmon anharmonicity can be as low as $-\alpha / 2 \pi \approx \qty{50}{\mega\hertz}$~\cite{mori2025highpowerreadouttransmonqubit, PhysRevApplied.23.034046}, while the inter-cavity coupling between microwave resonators can reach $J / 2 \pi \approx \qty{200}{\mega\hertz}$~\cite{deGraaf2025}, yielding a ratio $-\alpha/J \approx 1/4$. Our choice of $-\alpha/J = 1/3$ in \figpanel{fig3}{e} is therefore conservative. Furthermore, a coupling strength of $g / 2 \pi = 0.05J / 2 \pi \approx \qty{10}{\mega\hertz}$ is well within reach of current superconducting circuit technology~\cite{Kannan2020, Jouanny2025a}. The required strength of the central coupling point with $0 \leq \zeta < 2$ is modest and experimentally accessible~\cite{Kannan2020, Jouanny2025a}. A conservative estimation of typical decay rates for transmons and microwave resonators are $\Gamma_q = \qty{0.02}{\mega\hertz}$ and $\Gamma_c = \qty{0.1}{\mega\hertz}$, respectively (for resonators operating at $\qty{5}{\giga\hertz}$; even lower rates are achievable at smaller resonator frequencies)~\cite{annurev_qubits, Place2021, Somoroff2023, Kim2023, Biznarova2023, Kono2024, Bal2024, Acharya2025, PoorgholamKhanjari2025}. For $J / 2\pi = \qty{200}{\mega\hertz}$, these values correspond to $\Gamma_q = 1.6 \times 10^{-5}J$ and $\Gamma_c = 8 \times 10^{-5}J$, which we adopt in our analysis. 

For different values of $g$, the optimal process fidelity $\mathcal{F}_{\text{CZ,max}}$ and the corresponding gate time $\tau$ are shown in \figpanels{fig_fidelity}{c}{d}. For large $g$, the gate time $\tau$ is short, and decoherence from qubits and cavities does not constitute the primary limitation to fidelity~\cite{Abad2022, Abad2025}. In contrast, for small $g$, the increased gate time makes intrinsic decay the dominant source of infidelity. Consequently, the optimal performance is obtained at $g = 0.05J$ with gate time $\tau=297/J\approx\qty{236}{\nano\second}$, yielding a process fidelity of $\mathcal{F}_{\text{CZ}} \approx \qty{97.1}{\percent}$, corresponding to an average gate fidelity of $F_{\text{CZ}} \approx \qty{97.7}{\percent}$.


\section{Conclusion}
\label{sec4}

We have demonstrated how a controlled-Z (CZ) gate can be implemented using giant atoms (GAs) coupled to a structured waveguide. Starting from a minimal two-point coupling configuration, we showed that although the required decoherence-free interaction (DFI) can be realized, strong non-Markovian effects significantly limit the achievable gate fidelity. To address this limitation, we proposed an extended configuration that includes a third, intermediate coupling point. This modification shifts the decoherence-free frequencies closer to the center of the photonic band, thereby mitigating non-Markovian effects. With this design, we can achieve CZ gates with fidelities up to \qty{97.7}{\percent} when including realistic imperfections like typical resonator and transmon decay rates; remaining non-Markovian effects from spacing between coupling points also contribute to reducing the gate fidelity. 

Beyond the specific case of the CZ gate, our results establish a general strategy for engineering multi-photon processes in structured waveguide QED by tailoring interference conditions through multi-point coupling. This approach opens new possibilities for realizing other entangling operations, exploiting strongly correlated photonic dynamics, and extending gate protocols to larger networks of GAs. Moreover, the protocol is compatible with state-of-the-art superconducting circuits, where transmon coherence times are orders of magnitude longer than the gate duration, suggesting that experimental realization is within reach.

Looking ahead, our findings point toward several promising directions. One is the integration of multiple GAs in extended resonator arrays, enabling scalable architectures for quantum computation and simulation, such as passive-leakage reset for high-fidelity gates~\cite{thorbeck2024highfidelitygatestransmonusing}. Another is the exploration of richer structured environments, such as topological or chiral waveguides, where decoherence-free conditions could be further stabilized, or new types of protected interactions could emerge. More broadly, the interplay between photon–photon interactions~\cite{wang2024nonlinearchiralquantumoptics}, engineered dispersion, and non-Markovian dynamics in these systems defines a fertile platform for advancing fundamental studies of open quantum systems.


\begin{acknowledgements}
    
All code used in this work are open access via GitHub: \hyperlink{https://github.com/Q-WAVES/GiantAtoms/}{https://github.com/Q-WAVES/GiantAtoms/}. A.S. acknowledges fruitful discussions with Simon Pettersson Fors. G.C.~is supported by European Union's Horizon Europe programme HORIZON-MSCA-2023-PF-01-01 via the project 101146565 (SING-ATOM). W.R., A.S., and A.F.K.~acknowledge support from the Swedish Foundation for Strategic Research (grant number FFL21-0279). A.F.K.~is also supported by the Swedish Foundation for Strategic Research (grant number FUS21-0063), the Horizon Europe programme HORIZON-CL4-2022-QUANTUM-01-SGA via the project 101113946 OpenSuperQPlus100, and from the Knut and Alice Wallenberg Foundation through the Wallenberg Centre for Quantum Technology (WACQT).

\end{acknowledgements}


\bibliography{main,GA,NH_bib,BIC,doublon,footnote}

\end{document}